\documentclass[10pt, journal, twocolumn, letterpaper]{IEEEtran}

\usepackage{amssymb}
\usepackage{amsmath}
\usepackage{cite}
\newtheorem{theorem}{Theorem}
\newtheorem{lemma}[theorem]{Lemma}
\newtheorem{corollary}[theorem]{Corollary}
\newtheorem{definition}[theorem]{Definition}

\newcommand*{\tu}{two-universal}
\newcommand*{\tuty}{\tu ity}

\newcommand*{\DF}[2]{\mathcal{F}{(#1 \rightarrow #2)}} 
\newcommand*{\RF}[2]{\mathcal{R}{(#1 \rightarrow #2)}} 

\newcommand*{\fractext}[2]{#1/#2}

\newcommand*{\spc}{\,}
 
\newcommand*{\Pguess}{P_{\mathrm{guess}}}
\newcommand*{\Pav}[4]{\bar{P}_{#1}^{#2}(#3, #4)} 
\newcommand*{\binarysuc}[3]{\bar{P}_{#1}^{\max}(#2,#3)} 

\newcommand*{\devcl}[1]{\mathbf{C}^{2^{#1}}} 
\newcommand*{\devq}[1]{\mathbf{Q}^{#1}} 

\newcommand*{\Ps}{\mathcal{P}} 
\newcommand*{\Gs}{\mathcal{S}} 
\newcommand*{\POVM}{\mathrm{POVM}}
\newcommand*{\id}{\mathrm{id}} 

\newcommand*{\cFbin}[1]{\cF_{\mathrm{bin}}^{#1}} 
\newcommand*{\cFbal}[1]{\cF_{\mathrm{bal}}^{#1}} 

\newcommand*{\event}{\mathcal{E}} 

\newcommand*{\ket}[1]{| #1 \rangle}
\newcommand*{\bra}[1]{\langle #1 |}
\newcommand*{\spr}[2]{\langle #1 | #2 \rangle}
\newcommand*{\project}[1]{\ket{#1}\bra{#1}} 
\newcommand*{\tr}{\mathrm{tr}}

\newcommand*{\sfunc}{\varphi} 

\newcommand*{\kW}{\mathbf{W}} 
\newcommand*{\kV}{\mathbf{V}} 

\newcommand*{\staternd}{S} 
\newcommand*{\smallstaternd}{s}
\newcommand*{\statespace}{\cS}

\newcommand*{\keyrnd}{K} 
\newcommand*{\smallkeyrnd}{k}

\newcommand*{\channeloutput}[2]{#1_#2}

\newcommand*{\ent}{\mathcal{A}} 
\newcommand*{\comb}[2]{{#1};{#2}} 

\newcommand*{\cF}{\mathcal{F}} \newcommand*{\cG}{\mathcal{G}}
\newcommand*{\cH}{\mathcal{H}}

\newcommand*{\cS}{\mathcal{S}} \newcommand*{\cU}{\mathcal{U}}
 \newcommand*{\cW}{\mathcal{W}}
\newcommand*{\cX}{\mathcal{X}} \newcommand*{\cY}{\mathcal{Y}}
\newcommand*{\cZ}{\mathcal{Z}}

\newcommand*{\bp}{p}           
\newcommand*{\bbp}{\mathbf{p}} 

\newcommand*{\sbin}{\{0,1\}} 

\newcommand*{\cXf}[2]{\cX^{#2}_{#1}}
\newcommand*{\bQ}{\bar{Q}}
\newcommand*{\bk}{\bar{k}}

\DeclareMathOperator*{\ExpE}{\mathbb{E}}
\DeclareMathOperator*{\Prob}{\mathbb{P}}

\title{On the Power of Quantum Memory}

\author{Robert~K\"onig, Ueli~Maurer,~\IEEEmembership{Fellow,~IEEE,}
  and~Renato~Renner\thanks{The material in this paper was presented at
    the Seventh Workshop on Quantum Information Processing, Waterloo,
    Canada, January 2004. }\thanks{The authors are with the Department
    of Computer Science, Swiss Federal Institute of Technology (ETH)
    Z\"urich, CH-8092 Z\"urich, Switzerland (e-mail:
    rkoenig@inf.ethz.ch; maurer@inf.ethz.ch;
    renner@inf.ethz.ch).}\thanks{This work was partially supported by
    the Swiss National Science Foundation (SNF), project
    no.~200020-103847/1.}}

\begin{document}

\maketitle

\begin{abstract}
  
  We address the question whether quantum memory is more powerful than
  classical memory. In particular, we consider a setting where
  information about a random $n$-bit string $X$ is stored in
  $\smallstaternd$ classical or quantum bits, for $\smallstaternd <n$,
  i.e., the stored information is bound to be only partial. Later, a
  randomly chosen predicate $F$ about $X$ has to be guessed using only
  the stored information. The maximum probability of correctly
  guessing $F(X)$ is then compared for the cases where the storage
  device is classical or quantum mechanical, respectively.  We show
  that, despite the fact that the measurement of quantum bits can
  depend arbitrarily on the predicate $F$, the quantum advantage is
  negligible already for small values of the difference
  $n-\smallstaternd$.  Our setting generalizes the setting of Ambainis
  et al. who considered the problem of guessing an arbitrary bit
  (i.e., one of the $n$ bits) of $X$.
  
  An implication for cryptography is that privacy amplification by
  universal hashing remains essentially equally secure when the
  adversary's memory is allowed to be quantum rather than only
  classical. Since privacy amplification is a main ingredient of many
  quantum key distribution (QKD) protocols, our result can be used to
  prove the security of QKD in a generic way.
\end{abstract}


\section{Introduction}
\label{sec_intro}

It is a well-known fact that in $\smallstaternd$ quantum bits one
cannot reliably store more than $\smallstaternd$ classical bits of
information.\footnote{This is a direct consequence of the Holevo
  bound~\cite{Holevo73} stating that the accessible information
  contained in a quantum state cannot be larger than its von~Neumann
  entropy. This assertion is also a consequence of the general results
  proven in this paper (cf.\ Section~\ref{sec:binclbound}).}  In other
words, the raw storage capacity (like the raw transmission capacity)
of a quantum bit is just one bit of information. However, since
quantum memory can be read by an arbitrary measurement determined only
at the time of reading the memory, quantum memory can be expected to
be more powerful than classical memory in any context where a string
$X$ of $n>\smallstaternd$ bits of information is given (and hence can
be stored only partially) and it is determined only later which
information about $X$ is of interest.\footnote{A typical example of
  such a setting is the bounded-storage
  model~\cite{DziMau04,Vadhan03}.}

The simplest setting one can consider is that one must use the stored
information to guess $F(X)$ for a randomly chosen predicate $F: \cX
\rightarrow \sbin$.  Ambainis, Nayak, Ta-Shma, and
Vazirani~\cite{ANTV99,Nayak99} were the first to study such a setting
for the special case where $X$ is an $n$-bit string and $F(X)$ is an
actual bit (i.e., one of the $n$ bits) of $X$. Because in the quantum
case one can let the measurement of the stored quantum bits depend
arbitrarily on $F$, while in the classical case one can only read the
stored information, quantum memory is potentially more powerful.
However, we prove that having information about $X$ stored in $\smallstaternd$
quantum instead of $\smallstaternd$ classical bits is essentially useless for
guessing $F(X)$, even for optimal quantum storage and measurement
strategies. This is in accordance with the results
in~\cite{ANTV99,Nayak99} as well as with recent results on
communication complexity (see e.g., \cite{JaRaSe02}) where the power
of classical and quantum communication is compared.

In a cryptographic context, our results can be applied to the security
analysis of cryptographic primitives in a context where an adversary
might hold quantum information. An important example is \emph{privacy
  amplification} introduced by Bennett, Brassard, and
Robert~\cite{BeBrRo88} (see also~\cite{BBCM95}) which is a protocol
between two parties, Alice and Bob. The goal is to turn a common
$n$-bit string $X$, about which an adversary Eve has some partial
information, into a highly secure $\smallkeyrnd$-bit key $\keyrnd$. This can be
achieved as follows: Alice and Bob publicly agree on a function $G:
\sbin^n \rightarrow \sbin^\smallkeyrnd$ chosen from a \tu{} class of hash
functions\footnote{See Section~\ref{sec:not} for a definition of
  \tuty{}.} and then compute $\keyrnd=G(X)$.\footnote{Equivalently, they can
  use an extractor~\cite{NisZuc96}.}  It has been shown that, if Eve's
information about $X$ consists of no more than $\smallstaternd$ classical bits, the
final key is secure as long as $\smallkeyrnd < n-\smallstaternd$.\footnote{More precisely, her
  information is exponentially small in $n-\smallstaternd-\smallkeyrnd$.}

Similar to the previously described setting, it seems to be a
potential advantage for the adversary to have available $\smallstaternd$ quantum
instead of $\smallstaternd$ classical bits of information about $X$ because she
later learns the function $G$ and can let her measurement of the $\smallstaternd$
quantum bits depend on $G$.  This may allow her to obtain more
information about the final key $\keyrnd$. We prove that this is not the
case, i.e., privacy amplification remains equally secure against
adversaries holding quantum information.

This has interesting implications for quantum key distribution (QKD):
In a QKD protocol, Alice and Bob first exchange quantum information
(e.g., polarized photons) to generate a raw key $X$ which is only
partially secure, i.e., Eve has some quantum information $\rho$ about
$X$. In a second (purely classical) phase, Alice and Bob apply privacy
amplification to generate the final secret key $\keyrnd$. Our result on the
security of privacy amplification thus reduces the problem of proving
the security of a QKD protocol to the problem of finding a bound on
the number of qubits needed to (reliably) store Eve's information
$\rho$.  In~\cite{ChReEk04}, this fact has been exploited to show the
security of a generic QKD protocol which, in particular, implies the
security of many known protocols such as BB84~\cite{BenBra84}.  This
simplifies and generalizes\footnote{Most known security proofs are
  restricted to one specific QKD protocol.}  known security proofs
(see e.g., \cite{Mayers01}) which are based on completely different
techniques. It also generalizes a proof by Ben-Or~\cite{BenOr02} which
is based on a similar idea using results from communication complexity
theory~\cite{ASTVW98}.

The paper is organized as follows. In Section~\ref{sec:modknow}, we
introduce a general framework for modeling and quantifying knowledge
and storage devices.  The framework is then used in
Section~\ref{sec:boolpred} to state and prove bounds on the success
probability when guessing a binary predicate $F$ of $X$ given
information about $X$ stored in a quantum storage device
(Section~\ref{sec:binqbound}). These are then compared to the
situation where the information about $X$ is purely classical
(Section~\ref{sec:binclbound}).  In Section~\ref{sec:genpred}, the
results are extended to non-binary functions which then allows for
proving the security of privacy amplification against quantum
adversaries (Section~\ref{sec:pa}).

\section{Preliminaries} \label{sec:prel}

\subsection{Notation} \label{sec:not}

Let $\DF{\cX}{\cY}$ be the set of functions with domain $\cX$ and
range $\cY$. The set $\DF{\cX}{\sbin}$ of binary functions with
domain $\cX$, in the following called \emph{predicates} on $\cX$, is
denoted as $\cFbin{\cX}$.  Similarly, $\cFbal{\cX}:=\{f \in
\cFbin{\cX} : |f^{-1}(\{0\})| = |f^{-1}(\{1\})|\}$ is the set of
\emph{balanced predicates} on $\cX$.

Throughout this paper, random variables are denoted by capital letters
(e.g., $X$), their range by corresponding calligraphic letters
($\cX$), and the values they take on by lower case letters ($x$). The
event that two random variables $X$ and $Y$ take on the same value is
denoted as $X = Y$. In contrast, we write $X \equiv Y$ if two random
variables $X$ and $Y$ are identical (i.e., if $X = Y$ always holds).
The \emph{expectation} $\ExpE_{x\leftarrow P_X}[f(x)]$ of a function
$f$ on the random variable $X$ is given by $ \sum_{x \in \cX} P_X(x) \spc f(x) $.

For a channel $C$ from $\statespace$ to $\cW$ and a random variable
$\staternd$ on $\statespace$, we denote by
$\channeloutput{C}{\staternd}$ the output of $C$ on input $\staternd$,
i.e., if the channel is defined by the conditional distributions
$P_{W|\staternd=\smallstaternd}$ for $\smallstaternd \in \statespace$,
the joint probability distribution of $\channeloutput{C}{\staternd}$
and $\staternd$ is given by $P_{\channeloutput{C}{\staternd}
  \staternd}(w,\smallstaternd)=P(\smallstaternd)P_{\channeloutput{C}{\staternd}|\staternd=\smallstaternd}(w)$
for all $(w,\smallstaternd)\in\cW\times\statespace$.


A \emph{random function} $G$ from $\cX$ to $\cY$ is a random variable
taking values from the set $\DF{\cX}{\cY}$ of functions mapping
elements from $\cX$ to $\cY$.  The set of random functions from $\cX$
to $\cY$ is denoted as $\RF{\cX}{\cY}$.  If $G \in \RF{\cX}{\cY}$ is
uniformly distributed over $\DF{\cX}{\cY}$, it is called a
\emph{uniform random function} from $\cX$ to $\cY$.  Similarly, a
\emph{(uniform) random predicate} $F$ on $\cX$ is a random function
with (uniform) distribution over the set $\cFbin{\cX}$, and a
\emph{(uniform) balanced random predicate} is (uniformly) distributed
over the set $\cFbal{\cX}$. In the sequel, we will only use random
functions which are independent of all other (previously defined)
random variables. 

A random function $G$ from $\cX$ to $\cY$ is called\footnote{In the
  literature, two-universality is usually defined for families $\cG$
  of functions: A family $\cG$ is called \tu\ if the random function
  $G$ with uniform distribution over $\cG$ is \tu. For our purposes,
  however, our more general definition is more convenient.}
\emph{\tu{}} if $\Prob_{g\leftarrow P_G}[g(x) = g(x')] \leq
\fractext{1}{|\cY|} $ holds for any distinct $x, x' \in \cX$.  In
particular, $G$ is \tu{} if, for any distinct $x, x' \in \cX$, the
random variables $G(x)$ and $G(x')$ are independent and uniformly
distributed.  For instance, a uniform random function from $\cX$ to
$\cY$ is \tu.  Non-trivial examples where the distribution of $G$ is
over a smaller set of function (thus requiring less randomness) can,
e.g., be found in~\cite{CarWeg79} and~\cite{WegCar81}.


\subsection{Distance from Uniform}

The \emph{variational distance} between two distributions $P$ and $P'$
over an alphabet $\cZ$ is defined as
\[
    \delta(P, P')
  :=
    \frac{1}{2} \sum_{z \in \cZ} \bigl| P(z) - P'(z) \bigr| \ .
\]
The variational distance $\delta(P, \bar{P})$ of a distribution $P$
from the uniform distribution $\bar{P}$ (over the same alphabet $\cZ$)
is of particular interest in cryptographic applications. We will use
the abbreviation $d(P)$ for this quantity and refer to it as the
\emph{distance of $P$ from uniform}. For the distance of the
distribution of a random variable $Z$ from uniform, we also write
$d(Z)$ instead of $d(P_Z)$, and, more generally, for any event
$\event$, $d(Z|\event) := d(P_{Z|\event})$. Note that $d$ is a convex
function, i.e., for two probability distributions $P$ and $P'$, and
$q,q'\in [0,1]$ with $q+q'=1$, we have $ d(q \spc P + q' \spc P') \leq
q \spc d(P) + q' \spc d(P') $.

The distance $d(Z)$ of a random variable $Z$ from uniform has a
natural interpretation: It equals the probability that $Z$ deviates
from a uniformly distributed random variable $\bar{Z}$, in the
following sense.

\begin{lemma} \label{lem:equiva}
For any probability distribution $P_Z$ on $\cZ$ there exists a channel
$P_{\bar{Z}|Z}$ such that $P_{\bar{Z}}$ is the uniform distribution on
$\cZ$ and $\Prob_{(z,\bar{z})\leftarrow P_{Z\bar{Z}}}[z=\bar{z}]=1-d(Z)$.

\end{lemma}

For two random variables $Z$ and $W$, the \emph{(expected) distance of
  $Z$ from uniform given $W$} is defined (cf.~\cite{DziMau04}) as the
expectation of the distance of $Z$ from uniform conditioned on $W$,
i.e., $ d(Z|W) := \ExpE_{w\leftarrow P_W}[d(P_{Z|W=w})] $. It follows
directly from the convexity of $d$ that $d(Z|W) \geq d(Z)$, and, more
generally, for an additional random variable $V$ and an event
$\event$, $ d(Z|W V, \event) \geq d(Z|V, \event) $.

\section{Modeling Knowledge and Storage} \label{sec:modknow}

\subsection{Knowledge and Guessing}

Let $Z$ be a random variable and let $\ent$ be an entity with
knowledge described by a random variable $W$ (jointly distributed with
$Z$ according to some distribution $P_{Z W}$). Intuitively, one would
say that $\ent$ \emph{knows nothing} about $Z$ if $Z$ is uniformly
distributed given $\ent$'s knowledge $W$, i.e., $P_{Z W} \equiv P_Z
\times P_W$ where $P_Z$ is the uniform distribution. The following
straightforward generalization of Lemma~\ref{lem:equiva} suggests that
the distance $d(Z|W)$ of $Z$ from uniform given $W$ can be interpreted
as the probability of deviating from this situation.

\begin{lemma} \label{lem:equivb}
  For any probability distribution $P_{WZ}$ on $\cW\times\cZ$ 
there exists a channel $P_{\bar{Z}|WZ}$ such that $P_{\bar{Z}}$ is the
uniform distribution on $\cZ$, $P_{\bar{Z}W}\equiv P_{\bar{Z}}\times
P_{W}$, and $\Prob_{(z,\bar{z})\leftarrow P_{Z\bar{Z}}}[z=\bar{z}]=1-d(Z|W)$.
\end{lemma}

This is of particular interest in cryptography, where, for instance,
$\ent$ is an adversary with knowledge $W$ and where one wants to use
$Z$ as a key. Typically, a cryptosystem based on a key $\bar{Z}$ is
secure when $\bar{Z}$ is uniformly distributed and independent of
$\ent$'s knowledge. The lemma implies that, with probability
$1-d(Z|W)$, $Z$ is equal to such a perfect key $\bar{Z}$.  This means
that any statement which is true for an ideal setting where $\bar{Z}$
is used as a key automatically holds, with probability at least
$1-d(Z|W)$, for a real setting where $Z$ is the key.

The distance from uniform $d(Z|W)$ is also a measure for the maximum
success probability $\Pguess(Z|W)$ of an entity $\ent$ knowing $W$
when trying to guess $Z$,
\[
  \Pguess(Z|W) 
:= 
\max_{C} \Prob_{(w,z)\leftarrow P_{WZ}}[\channeloutput{C}{w}=z]\ ,
\]
where the maximum is over all channels $C$ from $\cW$ to $\cZ$.\footnote{Recall that
    $\channeloutput{C}{W}$ denotes the output of the channel $C$ on input $W$.}

The following lemma is an immediate consequence of the simple fact
that the best strategy for guessing $Z$ given $W=w$ is to choose a
value $\hat{z}$ maximizing the probability $P_{Z|W}(\hat{z}|w)$.

\begin{lemma} \label{lem:guessb}
  Let $W$ and $Z$ be random variables. Then $ \Pguess(Z|W) \leq
  \frac{1}{|\cZ|} + d(Z|W) $ where equality holds if $Z$ is binary.
\end{lemma}

\subsection{Selectable Knowledge}

The characterization of knowledge about a random variable $Z$ held by an entity $\ent$ in terms of
a random variable $W$ is sufficient whenever this knowledge is fully
accessible, e.g., written down on a sheet of paper or stored in a
classical storage device. However, in a more general context $\ent$
might have an option as to which information she can obtain. For
example, if her information about $Z$ is encoded into the state $\rho$
of a quantum
system, she may select \emph{one} arbitrary measurement to ``read it
out''. Formally, every measurement corresponds to a channel $W$ from the
state space of the quantum system to the set of possible measurement outcomes.
The situation is thus completely characterized by the set of
measurements (that is, channels) $\kW$ and the joint distribution of
$Z$ and $\rho$.  This setting is
discussed in detail in Section~\ref{sec:qstor}. Another (more
artificial) example might be a storage unit which can hold two bits
$\staternd\equiv B_1B_2$, but which allows only to read out \emph{one} of these
bits, i.e., $\ent$ can read either the value $B_1$ or $B_2$. In this
case, the situation is described by the joint distribution of $Z$ and
$\staternd$ and the set of channels $\{p_1,p_2\}$, where
 channel $p_i$ maps $(b_1,b_2)$ to $b_i$ for $i=1,2$.
 To model these situations, it is useful to introduce the following
notion.

\begin{definition}
A \emph{selectable channel} $\kW$ on $\statespace$ with range $\cW$ is a set of
channels from $\statespace$ to $\cW$.
\end{definition}


Consider now a setting as described above,
i.e., there is a system which is in a state described by a random
variable $\staternd$ on $\statespace$, and an entity $\ent$ has
access to $\staternd$ by means of a channel $W$ from a  set $\kW$.
In the following, we say that an 
entity $\ent$ {\em has selectable
knowledge $\channeloutput{\kW}{\staternd}$}, meaning that $\ent$ can learn the value of exactly
\emph{one} arbitrarily chosen random variable
$\channeloutput{W}{\staternd}$ with $W\in\kW$.  The
knowledge of $\ent$ about a random variable $Z$ can then be quantified
by a natural generalization of the distance measure introduced above.

\begin{definition} \label{def:nonunisk}
  Let $\staternd$ and $Z$ be random variables and let $\kW$ be a
  selectable channel on the range of $\staternd$.  The \emph{distance
    of $Z$ from uniform given $\channeloutput{\kW}{\staternd}$}, is
  \[
    d(Z|\channeloutput{\kW}{\staternd}) := \max_{W \in \kW} d(Z|\channeloutput{W}{\staternd}) \ .
  \]
\end{definition}

The significance of this generalized definition of \emph{distance from
  uniform}, e.g., in cryptography, is implied by a straightforward
extension of Lemma~\ref{lem:equivb}.

\begin{lemma} \label{lem:equivc}
  Let $\staternd$ and $Z$ be random variables and let $\kW$ be a
  selectable channel on the range of $\staternd$.  Then for any choice
  of an element $W$ of $\kW$, there exists a random variable $\bar{Z}$
  defined by a channel $P_{\bar{Z}|W_\staternd Z}$, such that
  $P_{\bar{Z}}$ is the uniform distribution on $\cZ$,
  $P_{\channeloutput{W}{\staternd}\bar{Z}}\equiv
  P_{\channeloutput{W}{\staternd}}\times P_{\bar{Z}}$, and $
  \Prob_{(z,\bar{z})\leftarrow P_{\bar{Z}Z}}[\bar{z} = z] \geq 1 -
  d(Z|\channeloutput{\kW}{\staternd}) $.
\end{lemma}

Similarly, Lemma~\ref{lem:guessb} can be generalized to obtain a bound
for the maximum success probability of an entity $\ent$ with
selectable knowledge $\channeloutput{\kW}{\staternd}$ when guessing $Z$,
\[
  \Pguess(Z|\channeloutput{\kW}{\staternd})  := \max_{W \in \kW} \Pguess(Z|\channeloutput{W}{\staternd}) \ .
\]

\begin{lemma} \label{lem:guessc}
  Let $\staternd$ and $Z$ be random variables and let $\kW$ be a
  selectable channel on the range of $\staternd$.  Then $
  \Pguess(Z|\channeloutput{\kW}{\staternd}) \leq \frac{1}{|\cZ|} +
  d(Z|\channeloutput{\kW}{\staternd}) $, where equality holds if $Z$
  is binary.
\end{lemma}


Consider now a situation where the information about $Z$ of an entity
$\ent$ is described by both some selectable knowledge
$\channeloutput{\kW}{\staternd}$, and, additionally, a random variable
$U$ which she can use to choose an element from $\kW$.  More
precisely, she applies some channel $C=P_{W|U}$ from $\cU$ to $\kW$ to
the random variable $U$ and then chooses to learn
$\channeloutput{W}{\staternd}$ for the resulting $W\equiv
\channeloutput{C}{U}\in\kW$. We will then be interested in the maximal
distance of $Z$ from uniform resulting from an optimal strategy used
by $\ent$. Such an optimal strategy consists simply of
(deterministically) choosing some $W\in\kW$ which maximizes
$\ExpE_{w\leftarrow
  P_{\channeloutput{W}{\staternd}}}[d(P_{Z|\channeloutput{W}{\staternd}=w,U=u})]$,
given $U=u$. We thus introduce the following quantity.
\begin{definition}
  Let $\staternd$, $U$ and $Z$ be random variables and let $\kW$ be a
  selectable channel on the range of $\staternd$. The \emph{distance
    of Z from uniform given $\channeloutput{\kW}{\staternd}$ and $U$}
  is defined as
\begin{equation} \label{eq:combstrat}
    d(Z|\comb{\channeloutput{\kW}{\staternd}}{U})
  := 
    \ExpE_{u\leftarrow P_U}\bigl[\max_{W \in \kW} 
d(Z|\channeloutput{W}{\staternd},U=u)
\bigr] \ .
\end{equation}
\end{definition}
It is easy to see that 
\[
 d(Z|\comb{\channeloutput{\kW}{\staternd}}{U})=d(Z|\channeloutput{\kV}{{(\staternd,U)}})
\]
for some selectable channel $\kV$ on $\statespace\times\cU$ which
models the fact that $\ent$ can choose an arbitrary strategy. In
particular, Lemma~\ref{lem:equivc} and Lemma~\ref{lem:guessc} still
hold when $\channeloutput{\kW}{\staternd}$ is replaced by 
$\comb{\channeloutput{\kW}{\staternd}}{U}$, where 
$\Pguess(Z|\comb{\channeloutput{\kW}{\staternd}}{U})$ is defined as
the maximal probability of $\ent$ when guessing $Z$ in the situation
described above.

It is a direct consequence of the properties of the variational
distance that knowledge of an additional random variable $U$ can only
increase the distance from uniform given selectable knowledge.

\begin{lemma} \label{lem:dcondgen}
  Let $\staternd$, $U$ and $Z$ be random variables and let $\kW$ be a
  selectable channel on the domain of $\staternd$. Then 
  \[
    d(Z|\comb{\channeloutput{\kW}{\staternd}}{U})\geq d(Z|\channeloutput{\kW}{\staternd}) \ .
  \]
\end{lemma}




\subsection{Storage Devices}

A (physical) storage device is a physical system where the information
it contains is determined by its physical state $\smallstaternd$.  Information is
stored in the device by choosing a state $\smallstaternd$ from its state space
$\statespace$. A storage device might provide different mechanisms to read out
this information, each of them resulting in some (generally only
partial) information about its state $\smallstaternd$. However, any possible
strategy of accessing the stored information can be described as a
channel mapping the memory state to a random variable $W$. We thus
define a \emph{storage device} with \emph{state space} $\statespace$ and range
$\cW$ as a selectable channel $\bbp$ from $\statespace$ to $\cW$.

As an example, consider the (artificial) storage device mentioned
above which allows to store two bits, but where only one of them can
be read out. Formally, this storage device is a selectable channel
$\bbp = \{\bp_1, \bp_2\}$ from the state space $\statespace = \sbin \times
\sbin$ to the set $\sbin$ where $\bp_m$ is the channel mapping
$(b_1, b_2)$ to $b_m$, for $m \in \{1, 2\}$.

The most trivial case is a classical storage device for storing $\smallstaternd$
bits and allowing to read out all $\smallstaternd$ bits without errors. Obviously,
its state $s$ can take one of $2^\smallstaternd$ possible values. Moreover, any
accessing strategy corresponds to a channel with input $s$. Formally,
a \emph{classical $\smallstaternd$-bit storage device} is defined as the selectable
channel $\devcl{\smallstaternd}$ containing all channels taking inputs
from the set $\sbin^\smallstaternd$. (In Section~\ref{sec:qstor}, we will give an
analogous definition for quantum storage devices.) Note that for a
random variable $Z$ and a
random variable $\staternd$  on $\sbin^\smallstaternd$, 
$d(Z|\channeloutput{\devcl{\smallstaternd}}{\staternd})=d(Z|\staternd)$. Thus we omit to mention the
selectable channel if it is clear from the context, e.g., we write
$d(Z|\comb{\staternd}{U})$ instead of
$d(Z|\comb{\channeloutput{\devcl{\smallstaternd}}{\staternd}}{U})$.

\subsection{Quantum Storage} \label{sec:qstor}

An $\smallstaternd$-qubit storage device is a quantum system of dimension $d=2^\smallstaternd$
where information is stored by encoding it into the state of the
system. This information can (partially) be read out by measuring the
system's state with respect to some (arbitrarily chosen) measurement
basis.  Each pure state of a $d$-dimensional quantum system corresponds to
a normalized vector $\ket{\psi}$ in a $d$-dimensional Hilbert space
$\cH_d$.  Equivalently, the set of pure states can be identified
with the set $\Ps(\cH_d) := \{\project{\psi} : \ket{\psi} \in \cH_d, |\spr{\psi}{\psi}| =
1\}$ where $\project{\psi}$ is the projection operator in $\cH_d$ along the
vector $\ket{\psi}$.  The set of all possible states of the quantum
system is then given by the set of mixed states $\Gs(\cH_d)$, which is
the convex hull of $\Ps(\cH_d)$.


It is well known from quantum information theory that the most general
strategy to access the information contained in a quantum system is to
perform a positive operator-valued measurement (POVM), which gives a
classical measurement outcome $W$. Any possible measurement is
specified by a family $\{E_w\}_{w \in \cW}$ of nonnegative operators
on $\cH_d$ satisfying $\sum_{w \in \cW} E_w = \id_{\cH_d}$. If the
system is in state $\rho$, the probability of obtaining the
(classical) measurement outcome $w \in \cW$ when applying measurement
$\{E_w\}_{w \in \cW}$ is given by $\bp_{\{E_w\}}(w|\rho) := \tr(E_w
\rho)$.

In the framework presented in the previous section, a
\emph{$d$-dimensional quantum storage device} $\devq{d}$ is thus
defined as the set of channels $\bp_{\{E_w\}}$ describing all possible
POVMs $\{E_w\}$ on a $d$-dimensional quantum state,
i.e.,
\[
    \devq{d} 
  := \
    \{ \bp_{\{E_w\}} : \{E_w\} \in \POVM(\cH_d) \} \ .
\]
A general way of describing this setting is to define the state
$\staternd$ of the storage device by a family of quantum states
$\{\rho_x\}_{x\in\cX}\subset\Gs(\cH_d)$, where $\rho_x$ is the
conditional state of the system given $X=x$, that is
$\staternd\equiv\rho_X$. Similar to the notation introduced for
classical storage devices $\devcl{\smallstaternd}$, we will also write
$\rho_X$ instead of $\channeloutput{\devq{d}}{{\rho_X}}$.

According to Definition~\ref{def:nonunisk}, the distance $d(Z|\rho_X)$
of a random variable $Z$ from uniform given $\rho_X$ can be written as
\[
d(Z|\rho_X) = \max_{\{E_w\}} d(Z|W)
\]
where the maximum is taken over all POVMs $\{E_w\}$ and where $W$ is
the measurement outcome of $\{E_w\}$ applied to the quantum state,
i.e., $P_{W|X=x}(w) = \tr(E_w \rho_x)$. Similarly, for an additional
random variable $U$,
\[
    d(Z|\comb{\rho_X}{U}) 
  = 
    \ExpE_{u \leftarrow P_U} \bigl[\max_{\{E_w^u\}} d(Z|W, U=u) \bigr]
\]
where, for each $u$, $\{E_w^u\}$ is a POVM and where $W$ is defined by
$P_{W|X=x, U=u}(w) = \tr(E_w^u \rho_x)$.

\section{Quantum Knowledge About Predicates} \label{sec:boolpred}

\subsection{The Quantum Binary Decision Problem}
\label{app:bindec}
We begin this section by stating a few known results about the
so-called quantum binary decision problem, which are central to the
proof of our main statements concerning quantum knowledge.

Let $\rho_0, \rho_1 \in \Gs(\cH)$ be arbitrary (mixed) states of a
quantum mechanical system $\cH$, and suppose that the system is
prepared either in the state $\rho = \rho_0$ or in $\rho = \rho_1$
with a priori probabilities $q$ and $1-q$, respectively.  The {\em
  quantum binary decision problem} is the problem of deciding between
these two possibilities by an appropriate measurement. Any decision
strategy can be summarized by a binary valued POVM $\{E_0,E_1\}$,
where the hypothesis $H_i: \rho=\rho_i$ is chosen whenever the outcome
is $i\in\sbin$. For a fixed strategy $\{E_0,E_1\}$, the probability
of choosing $H_i$, when the actual state is $\rho_j$, is given by
$\Prob[H_i|\rho=\rho_j]=\tr(E_i\rho_j)$, $i,j\in\sbin$.  Thus the
expected probability of success for this strategy equals
\[
  \Pav{q}{\{E_0,E_1\}}{\rho_0}{\rho_1} := q \spc \tr(E_0\rho_0)+(1-q) \spc \tr(E_1\rho_1) \ .
\]
The maximum achievable expected success probability in the binary
decision problem is the quantity
\[
    \binarysuc{q}{\rho_0}{\rho_1}
  :=
    \sup_{\{E_0,E_1\} \in \POVM}
      \Pav{q}{\{E_0,E_1\}}{\rho_0}{\rho_1} \ .
\]

The following theorem is due to Helstrom~\cite{Helstrom76}.  We state it
using the notation of Fuchs~\cite{Fuchs95} who also gave a simple proof
of it.

\begin{theorem} \label{thm:helstrom}
  Let $\rho_0, \rho_1 \in \Gs(\mathcal{H}_d)$ be two states, let $q
  \in [0,1]$, and let $\{\mu_i\}_{i=1}^d$ be the eigenvalues of the
  Hermitian operator
  $
    \Lambda :=q \spc \rho_0 - (1-q) \spc \rho_1
  $.
  Then the maximum achievable expected success probability in the
  quantum binary decision problem is
  \[
    \binarysuc{q}{\rho_0}{\rho_1}
  =
    \frac{1}{2} + \frac{1}{2} \sum_{i=1}^d |\mu_i| \ .
  \]
\end{theorem}

\subsection{Bounds on Quantum Knowledge} \label{sec:binqbound}

Let $X$ be a random variable and let $F$ be a randomly chosen
predicate on $\cX$. The goal of this section is to derive a bound on
the distance of $F(X)$ from uniform given knowledge about $X$ stored
in a quantum storage device.

Such knowledge is modeled by a family of quantum states
$\{\rho_x\}_{x\in\cX}$, where $\rho_x$ is the state of the quantum
system conditioned on the event that $X=x$. An explicit expression for
the corresponding quantity can be obtained using a result on the quantum binary
decision problem (cf.\ Section~\ref{app:bindec}).

\begin{lemma} \label{lem:eigenvaluessum}
  Let $X$ be a random variable with range $\cX$ and let $F$ be a
  random predicate on $\cX$.  Let
  $\{\rho_x\}_{x\in\cX}\subset\Gs(\cH_d)$ be a family of quantum
  states on a $d$-dimensional Hilbert space.  Then
  \[
      d(F(X)|\comb{\rho_X}{F})
    =
     \frac{1}{2} \ExpE_{f\leftarrow P_F} \Bigl[\sum_{j=1}^d |\mu^f_{j}|\Bigr] \ ,
  \]
  where $\{\mu^f_j\}_{j=1}^d$ are the eigenvalues of the Hermitian
  operator 
  \[
    \Lambda_{f} 
  := 
    \sum_{x: f(x) = 0} P_X(x) \rho_x
    - \sum_{x: f(x) = 1} P_X(x)\rho_x \ , \quad 
    \text{for $f \in \cFbin{\cX}$.}
  \]
\end{lemma}

\begin{proof}
  It suffices to show that
  \begin{equation} \label{eq:toprove}
    d(f(X)|\rho_X) = \frac{1}{2}\sum_{j=1}^d|\mu^f_j|
  \end{equation}
  for every $f \in \cFbin{\cX}$. Let thus $f$ be fixed and assume for
  simplicity that $P_{f(X)}(0) > 0$ and $P_{f(X)}(1) > 0$ (otherwise,
  (\ref{eq:toprove}) is trivially satisfied).
  
  Let $z \in \sbin$. Conditioned on the event that $f(X) = z$, the
  state $\rho$  equals $\rho_x$ with probability
  $P_{X|f(X)}(x|z)$. This situation can equivalently be described by
  saying that the system  is in the mixed state $\sigma^f_z \in
  \Gs(\cH_d)$, where
  \[
    \sigma^f_z = \sum_{x: f(x) = z} P_{X|f(X)}(x|z) \rho_x \ .
  \]
  The problem of guessing $f(X)$ thus corresponds exactly to the
  quantum binary decision problem described in
  Section~\ref{app:bindec}, i.e.,
  \[
      \Pguess(f(X)|\rho)
    =
      \binarysuc{P_{f(X)}(0)}{\sigma^f_0}{\sigma^f_1} \\
    =
      \frac{1}{2} + \frac{1}{2} \sum_{j=1}^d |\mu^f_j|
  \]
  where the second equality follows from Theorem~\ref{thm:helstrom}.
  Finally, since $f(X)$ is binary, equation~(\ref{eq:toprove}) follows
  from Lemma~\ref{lem:guessc}.
\end{proof}

The expression for the distance of $F(X)$ from uniform provided by
Lemma~\ref{lem:eigenvaluessum} is generally difficult to evaluate. The
following theorem gives a much simpler upper bound for this
quantity.\footnote{The main idea in the proof of
  Theorem~\ref{thm:mainschur} is to replace occurrences of density
  operators by their squares. The resulting expressions correspond to
  classical collision probabilities, as used in the well-known
  classical analysis of privacy amplification. The application of
  Jensen's inequality corresponds to the transition from the
  variational to the Euclidean distance. In this sense, this proof can
  be seen as a generalization of the classical derivation.  }

\begin{theorem} \label{thm:mainschur}
 Let $X$ be a random variable with range $\cX$ and let $F$ be a random
 predicate on $\cX$.  Let further 
$\{\rho_x\}_{x\in\cX}\subset\Gs(\cH_d)$ be a family of states on a
$d$-dimensional Hilbert space.  Then 
  \[
      d(F(X)| \comb{\rho_X}{F})
    \leq 
      \frac{1}{2} d^{\frac{1}{2}}
        \sqrt{\sum_{x,x' \in \cX} P_X(x) \spc P_X(x') \spc \lambda_{x,x'}
          \tr(\rho_x \rho_{x'})}
  \]
  where $\lambda_{x,x'} := 2 \spc \Prob_{f\leftarrow P_F} [f(x) = f(x')] - 1$, for $x,
  x' \in \cX$.
\end{theorem}

\begin{proof}
  We set out from the equation
  \[
      d(F(X)|\comb{\rho_X}{F})
    =
     \frac{1}{2} \ExpE_{f\leftarrow P_F}\Bigl[\sum_{j=1}^d |\mu^f_{j}|\Bigr]
  \]
  provided by Lemma~\ref{lem:eigenvaluessum}. Note that, for any $f
  \in \cFbin{\cX}$,
  \[
  \begin{split}
      \sum_{j=1}^d |\mu^f_{j}|
    \leq
      d^{\frac{1}{2}}\sqrt{\sum_{j=1}^d |\mu^f_{j}|^2}
    =
      d^{\frac{1}{2}}\sqrt{\tr(\Lambda_f^2)} \ ,
  \end{split}
  \]
  where the inequality is Jensen's inequality (applied to the convex
  mapping $x \mapsto x^2$) and where the equality is a consequence of
  Schur's (in)equality (cf.\ Lemma~\ref{lem:schur}), which can be
  applied because $\Lambda_f$ is Hermitian and thus also normal.  We
  conclude that
  \begin{equation}\label{eq:mainschur}
  \begin{split}
      d(F(X)|\comb{\rho_X}{F})
    & \leq
      \frac{1}{2} d^{\frac{1}{2}} \ExpE_{f\leftarrow P_F}\bigl[\sqrt{\tr(\Lambda_f^2)}\bigr] \\
     & \leq
      \frac{1}{2} d^{\frac{1}{2}}
        \sqrt{\ExpE_{f\leftarrow P_F} [\tr(\Lambda_f^2)]} \ ,
  \end{split}
  \end{equation}
  where Jensen's inequality is applied once again.

  By the definition of $\Lambda_f$ in Lemma~\ref{lem:eigenvaluessum},
  we have
  \[
  \begin{split}
      \tr(\Lambda_f^2)
    & =
        \sum_{\substack{x, x' \in \cX \\ f(x) = f(x')}}
          P_X(x) P_X(x')  \tr(\rho_x \rho_{x'}) \\
    & \quad
      - \sum_{\substack{x, x' \in \cX \\ f(x) \neq f(x')}}
          P_X(x) P_X(x')  \tr(\rho_x\rho_{x'}) \\
    & =
      \sum_{x, x' \in \cX}
        (2 \delta_{f(x), f(x')} - 1)
          P_X(x) P_X(x')  \tr(\rho_x \rho_{x'}) \ ,
  \end{split}
  \]
  where $\delta_{y,y'}$ is the Kronecker
  delta\footnote{$\delta_{y,y'}$ equals $1$ if $y=y'$ and
    $0$~otherwise.}.  The assertion then follows by taking the
  expectation of this expression over $F$ and combining the result
  with~(\ref{eq:mainschur}).
\end{proof}

If $F$ is \tu{}, the quantity on the right hand side of
Theorem~\ref{thm:mainschur} can be bounded by an expression which is
independent of the particular storage function.

\begin{corollary} \label{cor:binhash}
Let $X$ be a random variable with range $\cX$ and let $F$ be a \tu\
random predicate on $\cX$. Then for every family
$\{\rho_x\}_{x\in\cX}\subset\Gs(\cH_d)$ of states on a
$d$-dimensional Hilbert space
  \[
   d(F(X)|\comb{\rho_X}{F})
    \leq
      \frac{1}{2} d^{\frac{1}{2}}\sqrt{\sum_{x\in\mathcal{X}}P_X^2(x)} \ .
  \]
\end{corollary}

\begin{proof}
  Since $F$ is \tu{}, the values $\lambda_{x,x'}$ (as defined in
  Theorem~\ref{thm:mainschur}) cannot be positive for any distinct $x,
  x' \in \cX$. Since $\tr(\rho_x \rho_{x'})\geq 0$, we conclude that
  $\lambda_{x,x'} \spc \tr(\rho_x \rho_{x'}) \leq 0$ for $x
  \neq x'$.  Moreover, $\lambda_{x,x}=1$ and $\tr(\rho_x\rho_x) \leq 1$, for any $x \in \mathcal{X}$.  Combining these
  facts, the assertion follows directly from the upper bound given by
  Theorem~\ref{thm:mainschur}.
\end{proof}

Note that the expression under the square root is simply the collision
probability $P_C(X)$ of $X$. Hence, with the R\'enyi entropy
$R(X)=-\log_2 P_C(X)$, the above inequality can be rewritten as 
\begin{equation} \label{eq:allfunc}
  d(F(X)|\comb{\rho_X}{F})
  \leq
    \frac{1}{2} 2^{-\frac{R(X)-\smallstaternd}{2}} \ ,
\end{equation}
where $\smallstaternd$ is the number of qubits in which $X$ is stored, i.e.,
$\{\rho_x\}_{x\in\cX}\subset\Gs(\cH_{2^\smallstaternd})$. 

\subsection{Comparing Classical and Quantum Storage Devices} \label{sec:binclbound}

Since orthogonal states of a quantum system can always be perfectly
distinguished, a random variable $X$ can always be stored and
perfectly retrieved in a quantum storage device of dimension $d$ as
long as the size of the range of $X$ does not exceed $d$. Hence, a
classical $\smallstaternd$-bit storage device $\devcl{\smallstaternd}$
cannot be more powerful than a storage device 
$\devq{2^\smallstaternd}$ consisting of $\smallstaternd$ qubits. Formally, this can be stated as follows. For
any random variables $X$ and  $\staternd$ on  $\cX$
and $\sbin^\smallstaternd$, respectively, there is a family of states
$\{\rho_x\}_{x\in\cX}\subset\Gs(\cH_{2^\smallstaternd})$ such that
\begin{equation} \label{eq:compclassquant}
d(F(X)|\staternd F)\leq d(F(X)|\comb{\rho_X}{F})
\ , \quad \text{for any $F \in \RF{\cX}{\cY}$.}
\end{equation}

The following lemma shows that, on the other hand, a quantum storage
device can indeed be more useful than a corresponding classical
storage device.  However, we will see later that this is only true for
special cases, e.g., if the difference between the number $n$ of bits
to be stored and the capacity $\smallstaternd$ of the storage device
is small.



\begin{lemma} \label{lem:compex}
Let $X$ be uniformly distributed over $\sbin^2$ and let $F$ be a
uniform balanced predicate on $\sbin^2$.
Then for any random variable $\staternd$ on $\sbin$ defined by a channel
$P_{\staternd|X}$, 
  \[
   d(F(X)|\staternd F)
    \leq
      \frac{1}{4} \ .
  \]
Similarly, for every family
$\{\rho_x\}_{x\in\sbin^2}\subset\Gs(\cH_2)$ of quantum states on a
$2$-dimensional Hilbert space 
  \[
  d(F(X)|\comb{\rho_X}{F})
    \leq
      \frac{1}{2\sqrt{3}}\approx 0.289 \ ,
  \]
and there exists families
$\{\rho_x\}_{x\in\sbin^2}\subset\Gs(\cH_2)$ saturating this bound.
\end{lemma}

\begin{proof}
By the convexity of the variational distance, it suffices to consider
random variables $\staternd$ which depend in a deterministic way on $X$, that
is, $\staternd\equiv\sfunc_c(X)$ for some function $\sfunc_c:\sbin^2\rightarrow\sbin$.
  It can easily be verified (by an explicit calculation) that
  \[
  d(F(X)|\sfunc_c(X)F)\leq\frac{1}{4}
  \]
  for 
any function $\sfunc_c$ from $\sbin^2$
  to $\sbin$, and that equality holds for $\sfunc_c: (x_1,x_2)
  \mapsto x_1 \cdot x_2$ (i.e., $\sfunc_c(x_1,x_2) = 1$ if and only if
  $x_1=x_2=1$).  This proves the first (classical) statement of the
  lemma.

  For the second (quantum) statement, for the same reason as above, it
  suffices to consider pure states only. Let
$\{\project{\psi_x}\}_{x\in\sbin^2}\subset\Gs(\cH_2)$ 
be an arbitrary family of
pure quantum states.
 It follows
  from the linearity of the trace and Lemma~\ref{lem:quantJens},
  applied to the Hermitian operator $A := \sum_{x \in \cX}  \project{\psi_x}$, that
  \[
   \sum_{x,x'\in \cX} |\spr{\psi_x}{\psi_{x'}}|^2 \geq |\cX|^2/d \ .
  \]
The bound 
$d(F(X)|\comb{\project{\psi_X}}{F})\leq 1/(2\sqrt{3})$
can then be obtained
  from Theorem~\ref{thm:mainschur} with $\Prob_{f\leftarrow P_F}[f(x) = f(x')] =
  \frac{1}{3}$ for distinct $x, x'$ (implying $\lambda_{x,x'} = -\frac{1}{3}$).

  It remains to be proven that
  $d(F(X)|\comb{\project{\psi_X}}{F}) = 1/(2\sqrt{3})$
  for a family of states $\{\project{\psi_x}\}_{x\in\sbin^2}\subset\Gs(\cH_2)$. Such
  states can be defined by setting
  $\ket{\psi_{00}}$,$\ket{\psi_{01}}$
$\ket{\psi_{10}}$ and $\ket{\psi_{11}}$ to the vertices
  of a tetrahedron in $\Ps(\cH_2)$ (or, more precisely, in the Bloch
  sphere which corresponds to $\Ps(\cH_2)$). The assertion then
  follows from a straightforward calculation.
\end{proof}

Together with Lemma~\ref{lem:guessc}, Lemma~\ref{lem:compex} implies
that the maximum probability of correctly guessing a randomly chosen
balanced predicate $F$ about a random $2$-bit string $X$ is larger if
information about $X$ can be stored in one qubit ($P_q=0.789$) than if
this information is stored in one classical bit ($P_c=0.75$). Note
that this is in accordance with earlier results showing that one
individual qubit can be stronger than one classical bit (see, e.g.,
\cite{ANTV99}).

Surprisingly, this advantage of a quantum storage device becomes
negligible if the difference $n-\smallstaternd$ between the length $n$
of the bitstring $X$ and the number $\smallstaternd$ of bits/qubits of
the storage device becomes large. To see this, let us first state a
lower bound for the distance of $F(X)$ from uniform given the
knowledge stored in a classical storage device.

\begin{lemma} \label{lem:classical}
Let $X$ be uniformly distributed on $\sbin^n$ and  
let $F$ be a uniform random predicate on $\sbin^n$.
Then for any $\smallstaternd<n$ there exists a random variable $\staternd$ on $\sbin^\smallstaternd$ defined by
a channel $P_{\staternd|X}$ such that
  \begin{equation}
\label{eq:classlow}
      \frac{1}{2} C(2^{n-\smallstaternd}) 
    \leq
      d(F(X)|\staternd F)
  \end{equation}
  where $C(m) := \binom{m}{m/2} 2^{-m} = \sqrt{\frac{2}{\pi
      m}}(1+O(\frac{1}{m}))$. In particular,
\[
  \frac{1}{\sqrt{2 \pi}} 2^{-\frac{n-\smallstaternd}{2}}(1+O(2^{-(n-\smallstaternd)}))
    \leq
    d(F(X)|\staternd F) \ .
\]
\end{lemma}

\begin{proof}
  Let $\sfunc$ be a function from $\sbin^n$ to $\sbin^\smallstaternd$ such
  that for any $w \in \sbin^\smallstaternd$, the set $\sfunc^{-1}(\{w\}) := \{x \in
  \sbin^n : \sfunc(x) =w\}$ has size $2^{n-\smallstaternd}$. We claim that 
$S\equiv\sfunc(X)$ satisfies~\eqref{eq:classlow}.

For any fixed $w \in \sbin^\smallstaternd$ and $f \in
\cFbin{\sbin^n}$,
  \[
  \begin{split}
      d(f(X)|\sfunc(X)=w)
    & =
      \bigl| \Prob_{f\leftarrow P_F}[f(X) = 0|\sfunc(X)=w] - {\textstyle \frac{1}{2}} \bigr| \\
    & =
      \Bigl| \frac{k_f}{2^{n-\smallstaternd}} - {\textstyle \frac{1}{2}} \Bigr| \ ,
  \end{split}
  \]
  where $k_f := |f^{-1}(\{0\}) \cap \sfunc^{-1}(\{w\})|$.
    Since $F$ is uniformly distributed on the set $\cFbin{\sbin^n}$, we
  have $\Prob_{f\leftarrow P_F}[k_f=k] = \binom{2^{n-\smallstaternd}}{k}2^{-2^{n-\smallstaternd}}$ for $k \in
  \{0,\ldots,2^{n-\smallstaternd}\}$, hence
  \[
  \begin{split}
       d(f(X)|\sfunc(X)=w) 
     & =
      \sum_{k=0}^{2^{n-\smallstaternd}}
        \Bigl|\frac{k}{2^{n-\smallstaternd}}-\frac{1}{2}\Bigr|
          \binom{2^{n-\smallstaternd}}{k} 2^{-2^{n-\smallstaternd}} \\
     & =
       \frac{1}{2} C(2^{n-\smallstaternd}) \ ,
  \end{split}
  \]
  where the last equality follows from equation~(\ref{eq:factsum0}) of
  Lemma~\ref{lem:factsum}.  As $w \in \sbin^\smallstaternd$ was arbitrary, this
  concludes the proof. (The approximation for $C(m)$ can be obtained
  from Lemma~\ref{lem:stir}.)
\end{proof}

Combining Lemma~\ref{lem:classical} with
inequalities~(\ref{eq:allfunc}) and~(\ref{eq:compclassquant}), we
conclude that the distance from uniform has the same asymptotic
behavior for the classical and the quantum case: The knowledge about
the predicate $F(X)$ decreases exponentially in the difference
$n-\smallstaternd$ between the length of the bitstring $X$ and the
size $\smallstaternd$ of the storage device.

More precisely, since, for $n-\smallstaternd \geq 1$,
\[
  \frac{1}{2} C(2^{n-\smallstaternd}) 
\geq
  \frac{1}{2} 2^{-\frac{n-(\smallstaternd-1)}{2}}
\]
it follows from Lemma~\ref{lem:classical} and~(\ref{eq:allfunc}) that
there exists a random variable $S$ on $\{0,1\}^{\smallstaternd}$
defined by a channel $P_{S|X}$ such that $d(F(X)|\staternd F) \geq
d(F(X)|\comb{\rho_X}{F})$ for any family of states
$\{\rho_x\}_{x\in\sbin^n}\subset\Gs(\cH_{2^{\smallstaternd-1}})$. This
means that storing information about $X$ in $\smallstaternd$ classical
bits instead of $\smallstaternd-1$ quantum bits allows to predict
$F(X)$ with a lower error probability.

\section{From the Binary to the Non-Binary Case} \label{sec:genpred}

\subsection{Relations Between Bounds on Knowledge} \label{sec:nonbin}

We start with a lemma bounding the distance of a random variable $X$
from uniform by the distance of a binary hash value $F(X)$ from
uniform where $F$ is a randomly chosen balanced predicate. This is
related to the Vazirani XOR~lemma (see e.g.,~\cite{Gold95}), which
gives a similar bound for the case where $F$ is chosen randomly from
the set of all linear functions.\footnote{\label{ftn:vazirani} The
  following version of Vazirani's XOR~lemma is proved
  in~\cite{elbaz03}: $d(X)\leq \sqrt{|\cX|}\sqrt{\ExpE_{\ell\leftarrow
      P_L}[d(\ell(X))^2]}$, where $P_L$ is the uniform distribution on
  the set of all non-zero linear functions from $\cX$ to $\sbin$.  }

\begin{lemma}[Hashing Lemma] \label{lem:rel}
  Let $X$ be a random variable with range $\cX$ and let $F$ be a
  uniform balanced random predicate on $\cX$. Then
  \[
    d(X) \leq \frac{3}{2} \spc \sqrt{|\cX|} \spc d(F(X)|F) \ .
  \]
\end{lemma}


\begin{proof}
  For any probability distribution $Q$ over $\cX$ and any $f \in
  \cFbal{\cX}$, let $d_f(Q) := d(f(X'))$ be the distance between the
  uniform distribution and the distribution of $f(X')$ where $X'$ is a
  random variable distributed according to $Q$. We have to show that
  \begin{equation} \label{eq:lem}
    d(Q) \leq \frac{3}{2} \spc \sqrt{|\cX|} \spc \ExpE_{f\leftarrow P_F}[d_f(Q)] \ ,
  \end{equation}
  for any distribution $Q$ over $\cX$. Defining the coefficients
  $a_x(Q) := Q(x) - \frac{1}{|\cX|}$, and the sets $\cX^+_Q := \{x \in
  \cX : a_x(Q) \geq 0\}$ and $\cX^-_Q := \cX - \cX^+_Q$, we obtain
  \begin{equation} \label{eq:dist}
      d(Q)
    =
      \sum_{x \in \cX^+_Q} a_x(Q)
    =
      - \sum_{x \in \cX^-_Q} a_x(Q)
  \end{equation}
  and, for any $f \in \cFbal{\cX}$ and $\cXf{f}{0}:=\{x \in \cX : f(x)
  = 0 \}$,
  \begin{equation} \label{eq:fdist}
    d_f(Q) = \bigl| \sum_{x \in \cXf{f}{0}} a_x(Q) \bigr| \ ,
  \end{equation}
  respectively.  Note that, since $d$ is convex, $d_f$ is convex as
  well and thus so is its expected value $\ExpE_{f\leftarrow
    P_F}[d_f(\cdot)]$ (i.e., the function defined by $Q \mapsto
  \ExpE_{f\leftarrow P_F}[d_f(Q)]$).
  
  Let us first show that inequality~(\ref{eq:lem}) holds for
  distributions $\bQ$ over $\cX$ where the probabilities only take two
  possible values, $|\bQ(\cX)| \leq 2$, i.e., there exist $a^+ \geq 0$
  and $a^- \leq 0$ such that $a_x(\bQ) = a^+$ for $x \in \cX^+_{\bQ}$
  and $a_x(\bQ) = a^-$ for $x \in \cX^-_{\bQ}$. Then the value
  $d_f(\bQ)$ in~(\ref{eq:fdist}) only depends on the number $k(f):=
  |\cXf{f}{0} \cap \cX^+_{\bQ}|$ of values $x \in \cX^+_{\bQ}$ for
  which $f(x) = 0$.
  
  To get some intuition, consider the case where $|\cX^+_{\bQ}| =
  \frac{1}{2} |\cX|$. Since $f$ is randomly chosen, the expected
  deviation of $k(f)$ from its average value $\frac{1}{4}|\cX| $ is
  proportional to $\sqrt{|\cX|}$.  Furthermore, $d_f(\bQ)$ is
  proportional to this deviation and $a^+$, and $a^+$ is proportional
  to $d(\bQ)$ and inverse proportional to $|\cX|$. Neglecting the
  constants, this already shows that~(\ref{eq:lem}) holds in this
  particular case.
  
  Proving the exact statement~(\ref{eq:lem}) requires a little bit
  more computation. For any predicate $f \in \cFbal{\cX}$, expression
  (\ref{eq:fdist}) reads
  \[
  \begin{split}
      d_f(\bQ)
    & =
      \bigl| \sum_{x \in \cXf{f}{0} \cap \cX^+_{\bQ}} a^+
      + \sum_{x \in \cXf{f}{0} \cap \cX^-_{\bQ}} a^-  \bigr| \\
    & =
      \bigl| k(f) \spc a^+ + (\frac{n}{2} - k(f)) \spc a^- \bigr|
  \end{split}
  \]
  where $n := |\cX|$. With $s := |\cX^+_{\bQ}|$,
  expression~(\ref{eq:dist}) implies
  \[
    a^+ = \frac{d(\bQ)}{s} \qquad \text{and}
    \qquad a^- = - \frac{d(\bQ)}{n-s} \ ,
  \]
  and hence
  \[
  \begin{split}
      d_f(\bQ)
    & =
      \bigl|d(\bQ)
       \bigl( k(f) (\frac{1}{s} + \frac{1}{n-s})
         - \frac{n}{2} \spc \frac{1}{n-s} \bigr) \bigr| \\
    & =
      d(\bQ) \spc \bigl| k(f) - \frac{s}{2} \bigr| \spc \frac{n}{s (n-s)} \ .
  \end{split}
  \]
  Consequently, for $Q = \bQ$, inequality (\ref{eq:lem}) is equivalent
  to
  \[
      \frac{1}{| \cFbal{\cX}|}  \spc \frac{n}{s (n-s)} \spc \sum_{f \in \cFbal{\cX}}
        |k(f) - \frac{s}{2}|
    \geq
      \frac{2}{3 \sqrt{n}} \ .
  \]
  Since the term in the sum over $\cFbal{\cX}$ only depends on
  $k(f)$, the sum can be replaced by a sum over $k$, i.e., we have to
  show that
  \begin{multline} \label{eq:Qbars}
      \frac{1}{\binom{n}{\frac{n}{2}}} \spc \frac{n}{s (n-s)}
      \sum_{k = \max(0,s-\frac{n}{2})}^{\min(s,\frac{n}{2})}
        \binom{s}{k} \spc \binom{n-s}{\frac{n}{2}-k}
        \spc |k - \frac{s}{2}| \\
    =
      \frac{(\frac{n}{2}!)^2 \spc s! \spc (n-s)! \spc n }
           {n! \spc s (n-s)} \spc S_{n,s}
    \geq
      \frac{2}{3 \sqrt{n}}
  \end{multline}
  with
  \[
      S_{n,s}
    =
        \sum_{k = \max(0,-\frac{n}{2}+s)}^{\min(s,\frac{n}{2})}
      \frac{|k-\frac{s}{2}|}
        {k! \spc (s-k)! \spc
         (\frac{n}{2}-s+k)! \spc (\frac{n}{2}-k)!}\ .
  \]
  The term $S_{n,s}$ has different analytic solutions depending on
  whether $s$ is even or odd. Let us first assume that $s$ is even.
  Replacing the summation index $k$ by $\bk = k -\frac{s}{2}$ and
  making use of the symmetry of the resulting terms with respect to
  the sign of $\bk$, we get
  \[
  \begin{split}
      S_{n,s}
    & =
      2 \sum_{\bk=0}^{\min(\frac{s}{2}, \frac{n-s}{2})}
        \frac{\bk}
          {(\frac{s}{2}+\bk)! \spc (\frac{s}{2}-\bk)! \spc
           (\frac{n-s}{2}+\bk)! \spc (\frac{n-s}{2}-\bk)!} \\
    & =
      \frac{s (n-s)}{2 n \spc (\frac{s}{2}!)^2 \spc (\frac{n-s}{2}!)^2} \ ,
  \end{split}
  \]
  where the second equality follows from equation~(\ref{eq:factsum1})
  of Lemma~\ref{lem:factsum} with $a = \frac{s}{2}$ and $b =
  \frac{n-s}{2}$. A straightforward calculation then shows that for
  fixed $n$ the minimum of the left hand side of the inequality
  in~(\ref{eq:Qbars}) is taken for $s$ as close as possible to
  $\frac{n}{2}$, i.e., $s = 2 \lfloor \frac{n}{4} \rfloor$ and $n-s =
  2 \lceil \frac{n}{4} \rceil$, that is
  \[
  \begin{split}
      \frac{(\frac{n}{2}!)^2 \spc s! \spc (n-s)! \spc n }
          {n! \spc s (n-s)} \spc S_{n,s}
    & \geq
      \frac{\frac{n}{2}!^2 \spc s!\spc (n-s)!}
           {2 \spc n! \spc (\frac{s}{2}!)^2 \spc (\frac{n-s}{2}!)^2} \\
    & \geq
      \frac{\frac{n}{2}!^2 \spc (2 \lfloor \frac{n}{4} \rfloor) !
                           \spc (2 \lceil \frac{n}{4} \rceil) !}
           {2 \spc n! \spc (\lfloor \frac{n}{4} \rfloor !)^2
                 \spc (\lceil \frac{n}{4} \rceil !)^2} \ .
  \end{split}
  \]
  Lemma~\ref{lem:stir} is then used to derive a lower bound for the
  term on the right hand side of this inequality, leading to
  \begin{multline*}
      \frac{(\frac{n}{2}!)^2 \spc s! \spc (n-s)! \spc n }
          {n! \spc s (n-s)} \spc S_{n,s} \\
    \geq
      \sqrt{\frac{2}{\pi n}}
      \spc e^{\frac{2}{6n+1} + \frac{1}{24\lfloor\frac{n}{4}\rfloor+1}
                             + \frac{1}{24\lceil\frac{n}{4}\rceil+1}
              - \frac{1}{12n} - \frac{1}{6\lfloor\frac{n}{4}\rfloor}
                              - \frac{1}{6\lceil\frac{n}{4}\rceil}}
    \geq \frac{2}{3 \sqrt{n}} \ ,
  \end{multline*}
  where the last inequality holds for $n \geq 6$.

  Similarly, for $s$ odd,
  applying equation~(\ref{eq:factsum2}) of Lemma~\ref{lem:factsum}
  with $a = \frac{s-1}{2}$ and $b = \frac{n-s-1}{2}$ leads to
  \begin{multline*}
      S_{n,s}
    =
      2
        \sum_{\bk=0}^{\min(a, b)} 
        \frac{|\bk+\frac{1}{2}|}
          {(a+\bk+1)!
            \spc (a-\bk)!
            \spc (b+\bk+1)!
            \spc (b-\bk)!} \\
    =
      \frac{2}{n \spc (\frac{s-1}{2}!)^2 \spc (\frac{n-s-1}{2}!)^2} \ ,
  \end{multline*}
  resulting in the same lower bound $\frac{2}{3 \sqrt{n}}$ for the
  left hand side of the inequality in~(\ref{eq:Qbars}) for $n \geq 8$.
  Moreover, an explicit calculation shows that (\ref{eq:Qbars}) also
  holds for $n=2$, $n=4$, and $n=6$ which concludes the proof of
  inequality~(\ref{eq:lem}) for $Q = \bQ$ with $|\bQ(\cX)| \leq 2$.

  Let now $Q$ be an arbitrary distribution on $\cX$ and let $\Gamma$
  be the set of permutations on $\cX$ with invariant sets $\cX^+_Q$
  and $\cX^-_Q$, i.e., $\gamma(\cX^+_Q) = \cX^+_Q$ and
  $\gamma(\cX^-_Q) = \cX^-_Q$, for $\gamma \in \Gamma$. Since $d(Q) =
  d(Q \circ \gamma)$ for $\gamma \in \Gamma$, we find that
  \[
    \bQ := \frac{1}{|\Gamma|} \sum_{\gamma \in \Gamma} Q \circ \gamma
  \]
  is a probability distribution satisfying $d(\bQ) = d(Q)$ and taking
  identical probabilities for all elements in $\cX^+_Q$ as well as for
  all elements in $\cX^-_Q$, i.e., $|\bQ(\cX)| \leq 2$.  Since
  inequality~(\ref{eq:lem}) is already proven for distributions of
  this form, we conclude
  \[
  \begin{split}
      d(Q)
    =
      d(\bQ)
    & \leq
      \frac{3}{2} \spc \sqrt{|\cX|} \spc \ExpE_{f\leftarrow P_F}[d_f(\bQ)] \\
    & \leq
      \frac{3}{2} \spc \sqrt{|\cX|} \spc \frac{1}{|\Gamma|}
        \sum_{\gamma \in \Gamma} \ExpE_{f\leftarrow P_F}[d_f(Q \circ \gamma )] \ ,
  \end{split}
  \]
  where the second inequality is a consequence of the convexity of
  $\ExpE_{f\leftarrow P_F}[d_f(\cdot)]$. Assertion~(\ref{eq:lem}) then follows from
  $d_f(Q\circ\gamma)=d_{f\circ\gamma^{-1}}(Q)$, for all $f \in
  \cFbal{\cX}$, $\gamma \in \Gamma$, and the fact that
  $F\circ\gamma^{-1}$ is a uniform balanced random predicate, i.e.,
  $\ExpE_{f\leftarrow P_F}[d_{f\circ\gamma^{-1}}(Q)] =
  \ExpE_{f\leftarrow P_F}[d_f(Q)]$.
\end{proof}

In order to apply the hashing lemma to generalize the results of the
previous section to the non-binary case, we need a relation between
binary random functions (i.e., random predicates) and non-binary
random functions.

\begin{lemma} \label{lem:binuniv}
  Let $G$ be a \tu{} random function from $\cX$ to $\cY$ and let $F$
  be a uniform balanced random predicate on $\cY$.  Then the random
  predicate $H:=F \circ G$ is \tu{}.
\end{lemma}

\begin{proof}
  For any distinct $x, x' \in \cX$,
  \begin{multline*}
      \Prob_{h\leftarrow P_H}[h(x) = h(x')] 
    =
      \Prob_{g\leftarrow P_G}[g(x) = g(x')] \\
      \qquad \quad + (1-\Prob_{g\leftarrow P_G}  [g(x) = g(x')]) \spc
          \hspace{-2em} \Prob_{\substack{ f\leftarrow P_F \\ g\leftarrow P_{G|G(x)\neq G(x')}}}\hspace{-2em}[f(g(x)) = f(g(x')) ] \ .
  \end{multline*}
  Note that $\Prob_{f\leftarrow P_F, g\leftarrow P_{G|G(x)\neq G(x')}}[f(g(x)) = f(g(x')) ]$ is the
  collision probability of the uniform balanced random predicate $F$,
  $\Prob_{f\leftarrow P_F}[f(y) = f(y')]$ (for distinct $y, y' \in \cY$), which can
  easily be computed,
  \[
      \Prob_{f\leftarrow P_F}[f(y) = f(y')]
    =
      \frac{|\cY|-2}{2 \spc (|\cY|-1)} \ .
  \]
  Since $G$ is \tu{}, i.e., $\Prob_{g\leftarrow P_G}[g(x) = g(x')] \leq
  \frac{1}{|\cY|}$, we have
  \begin{multline*}
      \Prob_{h\leftarrow P_H}[h(x) = h(x')] \\
    =
        \Prob_{g\leftarrow P_G}[g(x) = g(x')] \bigl(1 -
        \Prob_{f\leftarrow P_F}[f(y) = f(y')]\bigr) \\
      + \Prob_{f\leftarrow P_F}[f(y) = f(y')] \\
    \leq
      \frac{1}{|\cY|}
      + \bigl(1 - \frac{1}{|\cY|}\bigr) \Prob_{f\leftarrow P_F}[f(y) = f(y')] \\
    =
      \frac{1}{|\cY|}
      + \bigl(1 - \frac{1}{|\cY|}\bigr) \frac{|\cY|-2}{2 \spc (|\cY|-1)}
    =
      \frac{1}{2} \ ,
  \end{multline*}
  i.e., the random predicate $H$ is \tu{}.
\end{proof}

Combining Lemma~\ref{lem:rel} and Lemma~\ref{lem:binuniv} leads to a
relation between the distance from uniform of the outcomes of binary
and general (non-binary) \tu{} functions on a random variable $X$,
given some knowledge
$\channeloutput{\kW}{\staternd}$.\footnote{\label{ftn:vaziranisecond}
  Using the version of Vazirani's XOR-Lemma stated in
  Footnote~\ref{ftn:vazirani}, the constant $\frac{3}{2}$ in the
  bound~(\ref{eq:condthmhashconst}) of Theorem~\ref{thm:hash} can be
  eliminated by replacing condition~(\ref{eq:condthmhash}) by the
  stronger requirement $\sqrt{\ExpE_{h\leftarrow
      P_H}[d(h(X)|\channeloutput{\kW}{\staternd})^2]}\leq\varepsilon$.
}

\begin{theorem} \label{thm:hash}
  Let $X$ and $\staternd$ be random variables on $\cX$ and
  $\statespace$, respectively and let $\kW$ be a selectable channel on
  $\statespace$. If, for all \tu{} random predicates $H$ on $\cX$,
  \begin{equation}\label{eq:condthmhash}
    d(H(X)|\comb{\channeloutput{\kW}{\staternd}}{H}) \leq \varepsilon \ ,
  \end{equation}
  then, for all \tu{} random functions $G$ from $\cX$ to $\cY$, 
  \begin{equation}\label{eq:condthmhashconst}
    d(G(X)|\comb{\channeloutput{\kW}{\staternd}}{G}) \leq \frac{3}{2} \sqrt{|\cY|} \spc \varepsilon \ .
  \end{equation}
\end{theorem}

\begin{proof}
From Definition~(\ref{eq:combstrat}), we have
\[
d(G(X)|\comb{\channeloutput{\kW}{\staternd}}{G})=\ExpE_{g\leftarrow
  P_G}\bigl[\max_{W\in\kW} d(g(X)|\channeloutput{W}{\staternd})\bigr]
\]
The expression in the maximum can then be bounded using
Lemma~\ref{lem:rel}, that is
\[
\begin{split}
d(g(X)|\channeloutput{W}{\staternd}) 
 & \leq
    \frac{3}{2}\spc\sqrt{|\cY|}\spc d(F(g(X))|\channeloutput{W}{\staternd}F).
\end{split}
\]
This leads to
\[
\begin{split}
d(G(X)|\comb{\channeloutput{\kW}{\staternd}}{G})
&\leq
\frac{3}{2}\spc\sqrt{|\cY|}
\spc \ExpE_{g\leftarrow
  P_G}\bigl[\max_{W\in\kW} d(F(g(X))|\channeloutput{W}{\staternd}F)\bigr]\\
&\leq 
\frac{3}{2}\spc\sqrt{|\cY|}
\spc \ExpE_{\substack{f\leftarrow P_F \\g\leftarrow P_G}}\bigl[\max_{W\in\kW}
d(f(g(X))|\channeloutput{W}{\staternd})\bigr]\\
\end{split}
\]
Defining $H:=F\circ G$, we obtain
\[
\begin{split}
d(G(X)|\comb{\channeloutput{\kW}{\staternd}}{G})
&\leq \frac{3}{2}\spc\sqrt{|\cY|}
\spc \ExpE_{h\leftarrow P_H}
\bigl[\max_{W\in\kW}
d(h(X)|\channeloutput{W}{\staternd})\bigr]\\
&=\frac{3}{2}\spc\sqrt{|\cY|}
\spc d(H(X)|\comb{\channeloutput{\kW}{\staternd}}{H}).
\end{split}
\]
 Finally, Lemma~\ref{lem:binuniv} states that $H$ is a \tu{} random
  predicate on $\cX$, hence the assertion of the theorem follows.

\end{proof}

\subsection{Application: Privacy Amplification with a Quantum Adversary} \label{sec:pa}

Consider two parties, Alice and Bob, being connected by an authentic
but otherwise completely insecure communication channel. Assume that
they initially share a uniformly distributed $n$-bit key $X$ about
which an adversary Eve has some partial information, where the only
bound known on Eve's information is that it consists of no more than
$\smallstaternd$ bits. Privacy amplification, introduced by Bennett, Brassard, and
Robert~\cite{BeBrRo88}, is a method to transform $X$ into an almost
perfectly secure key $\keyrnd$. It has been shown that if Alice and Bob
publicly (by communication over the insecure channel) choose a \tu{}
random function $G$ mapping the $n$-bit string to an $\smallkeyrnd$-bit string
$\keyrnd=G(X)$, for $\smallkeyrnd$ smaller than $n-\smallstaternd$, then the resulting string $\keyrnd$ is
secure (i.e., Eve has virtually no information about $\keyrnd$). Note that
$n-\smallstaternd$ is roughly Eve's entropy about the initial string $X$, i.e.,
privacy amplification with \tu{} random functions is asymptotically
optimal with respect to the number of extractable key bits. In our
formalism, the possibility of privacy amplification by applying a
(\tu{}) random function $G$, as proved in~\cite{BeBrRo88} (a
simplified proof has been given in~\cite{BBCM95}), reads
\begin{equation} \label{eq:paclass}
d(G(X)|\staternd G)= O(2^{-\frac{n-\smallstaternd-\smallkeyrnd}{2}}) \
\end{equation}
for any random variable $\staternd$ on $\sbin^\smallstaternd$ defined by a channel $P_{\staternd|X}$.

Combining the results from the previous section, we obtain a similar
statement for the situation where Eve's knowledge about $X$ is stored
in $\smallstaternd$ quantum instead of $\smallstaternd$ classical
bits. More precisely, we can derive a bound on the distance of the
final key $\keyrnd\equiv G(X)$ from uniform, from an adversary's point
of view, where $G$ is a \tu{} random function applied to an initial
string $X$, assuming only that the adversary's knowledge about $X$ is
stored in a limited number $\smallstaternd$ of
qubits.\footnote{\label{ftn:vaziranithird} Note that this is an
  example illustrating the fact that a bound on the expected distance
  of a single bit $H(X)$ from uniform
  $d(H(X)|\comb{\channeloutput{\kW}{\staternd}}{H})$ suffices to
  derive bounds on the expected distance from uniform
  $d(G(X)|\comb{\channeloutput{\kW}{\staternd}}{G})$ of a long key
  $G(X)$ obtained by privacy amplification. In the case of quantum
  knowledge, however, it is possible to prove even stronger statements
  for the single-bit case, resulting in a strengthened version of
  Corollary~\ref{cor:binhash}, which gives a bound on a quantity
  similar to $d(H(X)|\comb{\channeloutput{\kW}{\staternd}}{H})$ .
  Using this and Footnote~\ref{ftn:vaziranisecond}, the
  constant~$\frac{3}{4}$ in Corollary~\ref{cor:pa} can be replaced
  by~$\frac{1}{2}$.  }

\begin{corollary} \label{cor:pa}
  Let $X$ be a random variable with range $\cX$ and R\'enyi entropy
  $R(X)=n$ and let $G$ be a \tu{} random function from $\cX$ to
  $\sbin^\smallkeyrnd$.  Then, for any family of states
$\{\rho_x\}_{x\in\cX}\subset\Gs(\cH_{2^\smallstaternd})$ 
  \[
       d(G(X)|\comb{\rho_X}{G})
    \leq
      \frac{3}{4} \spc 2^{-\frac{n-\smallstaternd-\smallkeyrnd}{2}} \ .
  \]
\end{corollary}

\begin{proof}
  Theorem~\ref{thm:hash} together with Corollary~\ref{cor:binhash} implies
  \[
      d(G(X)|\comb{\rho_X}{G})
    \leq
      \frac{3}{4} \sqrt{2^\smallkeyrnd \cdot 2^\smallstaternd
        \spc \sum_{x \in \cX} P_X^2(x)}
  \]
for any family of states $\{\rho_x\}_{x\in\cX}\subset\Gs(\cH_{2^\smallstaternd})$.
 The corollary then follows from the
  definition of the R\'enyi entropy (cf.\ remark after the proof of
  Corollary~\ref{cor:binhash}).
\end{proof}

We thus have a quantum analogue to (\ref{eq:paclass}), implying that
privacy amplification remains equally secure (with the same
parameters) if an adversary has quantum rather than only classical
bits to store her information. Note that a similar bound follows 
from~\cite{BenOr02} together with a result of~\cite{Nayak99}, 
for the case where $G$ is the inner product with a randomly chosen string.

This generalization of the security proof of privacy amplification
immediately extends a result by Csisz\'ar and K\"orner~\cite{CsiKor78}
(see also \cite{Maurer93}) to the quantum case. Consider a situation
where Alice and Bob share information described by $N$ independent
realizations of random variables $X$ and $Y$, respectively, and where
Eve has information described by realizations of a classical random
variable $Z$. The result of~\cite{CsiKor78} says that the number of
secret key bits that can be generated by one-way communication (from
Alice to Bob) over a public channel is at least (roughly) $N (I(X;Y) -
I(X;Z))$, for large $N$.  The protocol that Alice and Bob have to
apply consists of an error correction step followed by a privacy
amplification step using a two-universal random function.  If we now
consider a situation where Eve holds $\smallstaternd$ qubits of
quantum information about $X$, it follows immediately from
Corollary~\ref{cor:pa} that the same protocol can be used to generate
a secret key of length roughly $N (I(X;Y) - \smallstaternd)$.

In most QKD protocols, Alice encodes some classical information $X$
into the state of a quantum system and sends it to Bob. Upon receiving
this state, Bob applies a measurement, resulting in classical
information $Y$. After this step, the adversary might hold some
quantum information about $X$ and $Y$. The situation is thus
characterized by classical random variables $X$ and $Y$ together with
the quantum system of Eve, where the size of her system depends on the
error rate tolerated by the protocol (see \cite{ChReEk04}). Hence, the
generalization of the Csisz\'ar-K\"orner bound described above
directly gives an expression for the amount of key that can be
generated by the protocol. In particular, it proves that the security
holds against any type of attack (including coherent measurements on
Eve's whole quantum system).

\section{Conclusions and Open Problems}

It is a fundamental question whether $\smallstaternd$ quantum bits are
more powerful than $\smallstaternd$ classical bits in order to store
information about an $n$-bit value $X$ (for $n>\smallstaternd$). We
considered the problem of answering a randomly chosen question $F$
about $X$, given only the stored information about $X$. The
uncertainty about the answer $F(X)$ is then a measure for the
usefulness of the stored information. It can be quantified in terms of
the distance of $F(X)$ from uniform conditioned on the stored
information, which, for binary questions $F$, corresponds to the
advantage over $1/2$ of the success probability when guessing $F(X)$.
It turns out that when storing a bitstring $X$ of length $n=2$ bits,
one quantum bit can indeed be more useful than one classical bit (cf.\ 
Lemma~\ref{lem:compex}). However, for larger values of
$n-\smallstaternd$, the difference between classical and quantum
memory becomes inessential.\footnote{As shown in
  Section~\ref{sec:binclbound}, $\smallstaternd$ classical bits can be
  more useful than $\smallstaternd-1$ quantum bits.}

We have shown that this has interesting implications for cryptography.
In particular, privacy amplification by \tu{} hashing remains secure
even against adversaries holding quantum information (cf.\ 
Corollary~\ref{cor:pa}).  This also leads to conceptually simpler and
more general security proofs for quantum key distribution, where
privacy amplification is used for the classical post-processing of the
raw key (cf.\ \cite{ChReEk04,BenOr02}).

It is well-known that so-called \emph{strong extractors} \cite{NisZuc96} can be used
to do privacy amplification in the classical case. While \tu{} hashing
can be seen as special case of this, the converse generally does not
hold. It is an open problem whether strong extractors are
sufficient to generate a key which is secure against a quantum
adversary in general. 

\section*{Acknowledgment}

The authors thank Gilles Brassard, Nicolas Gisin, and Stefan Wolf for
many interesting discussions. We also thank the anonymous referees for
helpful comments, in particular for pointing out the relation to
Vazirani's XOR-Lemma (cf.~Footnotes~\ref{ftn:vazirani}--\ref{ftn:vaziranithird}).

\bibliographystyle{IEEEtran}


\appendix



\begin{lemma}[Schur's inequality] \label{lem:schur}
  Let $A$ be a linear operator on a $d$-dimensional Hilbert space
  $\cH_d$ and let $\{\mu_i\}_{i=1}^d$ be its eigenvalues. Then
  \[
      \sum_{i=1}^d |\mu_{i}|^2
    \leq
      \tr(A A^{\dagger}) \ ,
  \]
  with equality if and only if $A$ is normal (i.e., $A
  A^{\dagger}=A^{\dagger} A$).
\end{lemma}

\begin{proof}
  See, e.g., \cite{HorJoh85}.
\end{proof}
 
\begin{lemma} \label{lem:quantJens}  
  Let $A$ be a normal operator on a $d$-dimensional Hilbert space
  $\cH_d$. Then
  \[
    |\tr(A)|^2 \leq d \cdot \tr(A A^{\dagger}) \ .
  \]
\end{lemma}

\begin{proof}
  Since $A$ is normal, we have
  \[
    \tr(A) = \sum_{i=1}^d \mu_i
  \quad \text{and} \quad
    \tr(A A^{\dagger}) = \sum_{i=1}^d |\mu_i|^2 \ ,
  \]
  where $\{\mu_i\}_{i=1}^d$ are the eigenvalues of $A$. The assertion
  then follows from Jensen's inequality stating that
  \[  
    \Bigl|\sum_{i=1}^d \mu_i\Bigr|^2 
  \leq
    d \cdot \sum_{i=1}^d |\mu_i|^2 \ .
  \]
\end{proof}

\begin{lemma} \label{lem:factsum}
  Let $a,b \in \mathbb{N}$. Then the following equalities hold:
  \begin{equation}
      \sum_{z=0}^{2a} \label{eq:factsum0}
        \binom{2a}{z} \cdot \bigl| \frac{1}{2} - \frac{z}{2a} \bigr|
    = 
      \frac{1}{2} \binom{2a}{a} 
  \end{equation}
  \begin{equation}
  \begin{split} \label{eq:factsum1}
      \sum_{z = 0}^{\min(a,b)} \vspace{-0.5em}
        \frac{z}{(a+z)!\spc (a-z)!\spc (b+z)!\spc (b-z)!} 
    = 
      \frac{a b}{2 (a+b) \spc (a!)^2 \spc (b!)^2}  
  \end{split}
  \end{equation}
  \begin{equation}
  \begin{split}
      \sum_{z = 0}^{\min(a,b)} \label{eq:factsum2}
        \frac{z+\frac{1}{2}}{(a+z+1)!\spc (a-z)!\spc (b+z+1)!\spc (b-z)!} \\
     = 
      \frac{1}{2 (a+b+1) \spc (a!)^2 \spc (b!)^2} \ . \hspace{-2em}
  \end{split}
  \end{equation}
\end{lemma}

\begin{proof}
  The first equality follows from a straightforward calculation, using
  the identity $\binom{a}{z} \cdot \frac{z}{a} = \binom{a-1}{z-1}$.
  The second and the third equality can be obtained with Zeilberger's
  algorithm~\cite{Zeilberg90} which is implemented in many standard
  computer algebra systems (e.g., Mathematica or Maple).
\end{proof}

\begin{lemma}[Stirling's approximation] \label{lem:stir}
  For $n \in \mathbb{N}$,
  \[
      \sqrt{2 \pi} n^{n+\frac{1}{2}} e^{-n+\frac{1}{12 n + 1}}
    <
      n!
    < \sqrt{2 \pi} n^{n+\frac{1}{2}} e^{-n+\frac{1}{12 n}} \ .
  \]
\end{lemma}

\begin{proof}
  A proof of this extension of Stirling's approximation can be found
  in \cite{Feller68}.
\end{proof}

\end{document}